%% file: main.tex
\RequirePackage[l2tabu, orthodox]{nag}

\documentclass[sigconf]{acmart}

\AtBeginDocument{%
  \providecommand\BibTeX{{%
    \normalfont B\kern-0.5em{\scshape i\kern-0.25em b}\kern-0.8em\TeX}}}

\usepackage[T1]{fontenc}            
\usepackage[utf8]{inputenc}         
\usepackage{microtype}              
\usepackage{booktabs}
\usepackage[scaled=.8]{beramono}    
\usepackage[abbreviations]{foreign} 
\usepackage{balance}                
\usepackage{listings}
\usepackage{pifont}                 
\usepackage{multirow}
\usepackage{subcaption}
\usepackage[all]{nowidow}
\usepackage[multiple]{footmisc}
\usepackage{csquotes}
\usepackage{doi}
\usepackage{cleveref}               

\input{commands}

\setcopyright{none}
\copyrightyear{2024}
\acmYear{2024}
\acmDOI{10.1145/3643916.3644415}
\acmConference[ICPC '24]{32nd IEEE/ACM International Conference on Program Comprehension}{April 15--16, 2024}{Lisbon, Portugal}
\acmPrice{15.00}
\acmISBN{979-8-4007-0586-1/24/04}

\acmBooktitle{32nd IEEE/ACM International Conference on Program Comprehension (ICPC '24), April 15--16, 2024, Lisbon, Portugal}

\begin{document}

\title{Lightweight Syntactic API Usage Analysis with UCov}

\author{Gustave Monce}
\email{gustave.monce@labri.fr}
\affiliation{%
  \institution{Univ. Bordeaux, CNRS, Bordeaux INP, LaBRI, UMR 5800}
  \country{France}
}

\author{Thomas Couturou}
\email{thomas.couturou@labri.fr}
\affiliation{%
  \institution{Univ. Bordeaux, CNRS, Bordeaux INP, LaBRI, UMR 5800}
  \country{France}
}

\author{Yasmine Hamdaoui}
\email{yasmine.hamdaoui@labri.fr}
\affiliation{%
	\institution{Univ. Bordeaux, CNRS, Bordeaux INP, LaBRI, UMR 5800}
	\country{France}
}

\author{Thomas Degueule}
\email{thomas.degueule@labri.fr}
\affiliation{%
  \institution{Univ. Bordeaux, CNRS, Bordeaux INP, LaBRI, UMR 5800}
  \country{France}
}

\author{Jean-Rémy Falleri}
\email{falleri@labri.fr}
\affiliation{%
  \institution{Univ. Bordeaux, CNRS, Bordeaux INP, LaBRI, UMR 5800}
  \institution{Institut Universitaire de France}
  \country{France}
}


\begin{abstract}
Designing an effective API is essential for library developers as it is the lens through which clients will judge its usability and benefits, as well as the main friction point when the library evolves.
Despite its importance, defining the boundaries of an API is a challenging task, mainly due to the diverse mechanisms provided by programming languages that have non-trivial interplays.
In this paper, we present a novel conceptual framework designed to assist library maintainers in understanding the interactions allowed by their APIs via the use of \emph{syntactic usage models}.
These customizable models enable library maintainers to improve their design ahead of release, reducing friction during evolution.
The complementary \emph{syntactic usage footprints} and coverage scores, inferred from client code using the API (\eg documentation samples, tests, third-party clients), enable developers to understand in-the-wild uses of their APIs and to reflect on the adequacy of their tests and documentation.
We implement these models for Java libraries in a new tool \ucov and demonstrate its capabilities on three libraries exhibiting diverse styles of interaction:~\jsoup, \cli, and \spark.
Our exploratory case study shows that \ucov provides valuable information regarding API design and fine-grained analysis of client code to identify under-tested and under-documented library code.
\end{abstract}

\ccsdesc[500]{Software and its engineering~Software libraries and repositories}
\ccsdesc[500]{Software and its engineering~Software maintenance tools}
\ccsdesc[300]{Software and its engineering~Software design engineering}

\keywords{software library, API design, API usage, API documentation}


\maketitle

\input{intro}
\input{background}
\input{usagemodel}
\input{ucov}
\input{eval}

\input{discussion}
\input{rw}
\input{conclusion}

\begin{acks}
This work was partially funded by the French National Research Agency through grant ANR ALIEN (ANR--21--CE25--0007). We thank the anonymous reviewers for their precious comments.
\end{acks}

\balance
\bibliography{main}
\bibliographystyle{plainnat}

\end{document}

%% file: commands.tex
\definecolor{keywordscolor}{RGB}{0, 51, 179}
\definecolor{stringcolor}{RGB}{6, 125, 23}
\definecolor{commentscolor}{RGB}{140, 140, 140}
\definecolor{annotationscolor}{RGB}{100, 100, 100}
\definecolor{lstbgcolor}{RGB}{251, 251, 251}

\newcommand*{\ijava}[1]{\lstinline[basicstyle=\ttfamily,language=Java,keywordstyle=\mdseries\color{keywordscolor},morekeywords={sealed}]{#1}}
\newcommand*{\etclst}{\colorbox{blue!10}{\textbf{\vphantom{A}\dots}}}

\newcommand*{\ucov}{\textsc{UCov}\@\xspace}
\newcommand*{\jsoup}{\textsc{jsoup}\@\xspace}
\newcommand*{\cli}{\textsc{commons-cli}\@\xspace}
\newcommand*{\spark}{\textsc{spark}\@\xspace}

%% file: intro.tex
\section{Introduction}
Application Programming Interfaces (APIs) govern the interactions between a library and client projects using it.
In particular, they specify which elements of a library (\eg types, methods, fields) can be accessed from the outside and the valid interactions with them.
Designing an effective API is essential for library developers as it is the lens through which clients judge its usability and benefits and the main friction point when the library evolves~\cite{robillard2009makes, ochoa2022breaking}.
Despite its importance, defining the boundaries of an API is a challenging task, mainly due to the diverse mechanisms provided by programming languages and paradigms that either aim specifically at designing APIs or indirectly affect them as a side effect (\eg visibilities, subclass restriction and sealing, name binding rules, overloading and overriding, polymorphism).

Given the complex interplay among those mechanisms, it is difficult for library maintainers to maintain an exhaustive and accurate mental model of their API and the interactions it enables.
This complexity is compounded by the various \emph{styles} libraries can adopt (\eg frameworks and inversion of control, fluent interfaces) and the variety of ways that client code can interact with individual API elements.
For instance, methods may either be invoked directly or overridden, and non-abstract classes may be instantiated, extended, referenced, \etc.
This impairs the maintainers' ability to fully understand how their APIs are---or could be!---utilized in the wild.
As a result, libraries sometimes inadvertently allow some kinds of interactions that their developers did not intend.
This led to best API design practices such as ``minimize accessibility'' or ``forbid subclassing by default'' that are widespread in different communities (\eg in Java~\cite{bloch2008effective}).
To minimize the likelihood of bad outcomes on the client side, library developers should restrict the possible interactions with their APIs to those that are \emph{intended} and \emph{validated}.
Indeed, unintended interactions are a recipe for buggy client code and frustrated developers who may opt for an alternative library providing similar services with better support.
However, there is no lightweight, effective way of extracting and maintaining a transparent, comprehensive API model, nor is there a way to validate the extent to which the interactions with an API are validated, documented, and representative of actual uses.
Current approaches in the literature can successfully analyze the use of API symbols but do not account for the variety of possible interactions with an individual symbol~\cite{qiu2016understanding,harrand2022api}.

In this paper, we introduce a novel conceptual framework aimed at helping library maintainers better understand the boundaries of their APIs via the use of so-called \emph{syntactic usage models} (SUMs).
SUMs enable maintainers to reason about the interactions allowed by their APIs throughout their evolution.
Based on the SUMs, we define \emph{syntactic usage footprints} (SUFs), which measure the extent to which a given piece of code using the library (\eg third-party clients, documentation samples, library tests) exercises its API and covers its possible uses.
We show how different SUFs can be easily compared, enabling library maintainers to answer questions such as ``\textit{do our tests validate the interactions found in third-party clients?}'' or ``\textit{do the examples in our documentation align with in-the-wild uses of our API?}'' (\Cref{sec:models}).

We present \ucov (\emph{Usage COVerage}), an implementation of these models for the Java programming language that leverages static analysis to automatically infer them from Java source code (\Cref{sec:ucov}).
We illustrate the benefits of \ucov and the underlying models with an exploratory case study of three libraries implementing various styles of interaction:~\jsoup, a library for HTML manipulation;\footnote{\url{https://jsoup.org/}}~Apache \cli, a library for implementing command-line interfaces;\footnote{\url{https://commons.apache.org/proper/commons-cli/}} and \spark, a simple web framework\footnote{\url{https://sparkjava.com/}} (\Cref{sec:eval}).
From our experiments with \ucov, we lay out a research agenda for syntactic models and envision application scenarios in other areas (\Cref{sec:discussion}).
~\cite{monce_2024_10571867}

Overall, our syntactic models and \ucov provide a novel means for library maintainers to understand the extent of possible interactions allowed by their APIs and to oversee their use in client code.
By offering a robust conceptual framework for usage coverage and a first implementation for Java, we aim to empower library maintainers with the insights needed to improve their API's design, documentation, and tests, thereby reducing friction with client code and fostering long-term maintenance.

%% file: background.tex
\section{Background and Motivation}
\label{sec:background}
Understanding how client code interacts with APIs is instrumental in prioritizing development and documentation efforts.
Therefore, numerous studies have examined methods to extract usage data from client code to identify notable \emph{hotspots} and \emph{coldspots}, \ie parts of the APIs that are over- or under-utilized~\cite{stylos2009improving,sawant2017fine,thummalapenta2008spotweb}.
In a recent study, \citeauthor{qiu2016understanding} delved into the usage of Java's standard library and third-party libraries across a corpus of over 5,000 projects, encompassing 150M+ lines of code~\cite{qiu2016understanding}.
Various tools and studies adopt distinct methodologies for gathering usage data, whether from bytecode~\cite{harrand2022api}, source code and resolved ASTs~\cite{qiu2016understanding,lammel2011large, deroover2013multi}, or other sources~\cite{stylos2009improving}.
A unifying theme across these works and their underlying models of API usage is their focus on determining \emph{whether a specific API symbol is accessed} in client code, rather than exploring \emph{how it is used}.
This emphasis stems from the primary objectives of these studies:~assess the frequency, popularity, and coverage of API symbols in client code, rather than exploring the diverse \emph{uses} of these symbols.
Indeed, the way a symbol is used has no influence on its popularity. Extending, instantiating, or simply referencing a class all contribute equally to its popularity.

We argue that this approach to modeling library usage is too coarse-grained to allow library maintainers to understand fully the implications of the design of their API\@.
For instance, let us consider a library designer providing a simple \ijava{public class C} and a client instantiating it with \ijava{C c = new C()}.
Existing usage models would indicate that there is one exported symbol (\texttt{C}) and that this symbol is used in client code.
Based on this information, library maintainers could conclude that the design of their API is satisfactory.
However, this declaration does not prevent clients from subclassing the provided class with \ijava{class ClientC extends C}.
The aforementioned models do not provide any means to distinguish this particular use from other uses in client code.
This is a missed opportunity, as library maintainers could have reconsidered the API early on to prevent subclassing, disallowing clients from extending the class, and thus freeing themselves from supporting this scenario.
The approaches mentioned above are oblivious to this distinction, while our work emphasizes it.

There is a wide variety of literature addressing the problem of mining and recommending API usage patterns and protocols~\cite{zhong2009mapo,wang2013mining}.
These studies typically analyze library code and client code to infer automata and probabilistic graphs that represent legal sequences of API invocations and check whether client code complies with them.
While these approaches go beyond our objectives and consider some kind of semantic relation between API elements, they are unable to differentiate between the different interactions allowed for a specific element.
In contrast to previous works, our approach emphasizes the diverse ways a particular symbol may be utilized in client code, leveraging the \emph{syntactic usage models} introduced in the next section.

%% file: usagemodel.tex
\section{Syntactic API Usage}
\label{sec:models}

Our analysis of API usage is built upon two distinct models:~syntactic usage models, which are extracted from library code and represent the \emph{legal} (possible) uses of an API (\Cref{sec:sum}), and syntactic usage footprints, which are extracted from client code and represent the \emph{actual} uses of an API (\Cref{sec:footprint}).
From these two models, we derive various metrics, including a dedicated API coverage metric (\Cref{sec:metrics}).
These models are formulated independently of any specific programming language and can be adapted for diverse analysis scenarios.
However, we reference Java syntax and semantics throughout this section for illustrative purposes.

\begin{table*}[tb]
	\centering
	\small
	\caption{Syntactic usage footprints for the API of \Cref{lst:arraylist}}
	\label{tab:footprint}
	\begin{tabular}{lrll}
		\textbf{Client} & \textbf{Statement} & \textbf{Symbol} & \textbf{Use} \\
		\midrule
		\multirow{5}{*}{\Cref{lst:lib-snippet}} & \ijava{ArrayList<Integer> lst} & \ijava{public class ArrayList<E>} & \textit{Referenced} \\
		& \ijava{new ArrayList<Integer>()} & \ijava{public class ArrayList<E>} & \textit{Instantiated} \\
		& \ijava{new ArrayList<Integer>()} & \ijava{public ArrayList<E>()} & \textit{Invoked} \\
		& \ijava{lst.add(42)} & \ijava{public boolean add(E e)} & \textit{Invoked} \\
		& \ijava{lst.add(1337)} & \ijava{public boolean add(E e)} & \textit{Invoked} \\
		\midrule
		\midrule
		\multirow{2}{*}{\Cref{lst:fw-snippet}} & \ijava{class MyArrayList<E> extends ArrayList<E>} & \ijava{public class ArrayList<E>} & \textit{Extended} \\
		& \ijava{@Override public boolean add(E e)} & \ijava{public boolean add(E e)} & \textit{Overridden} \\
		\bottomrule
	\end{tabular}
\end{table*}

\subsection{Syntactic Usage Model (SUM)}
\label{sec:sum}
We consider that a library defines a set of \emph{symbols} $\mathcal{S}$ (\ie named entities), each of a particular \emph{kind} and with a \emph{declaration} that specifies its properties.
For instance, the Java standard library defines the symbol \ijava{public class ArrayList<E>} of kind \emph{Class}, the symbol \ijava{public final static PrintStream out} of kind \emph{Field}, and the symbol \ijava{public void println()} of kind \emph{Method}.
These symbols can be exported, allowing client code to access them.
Together, the exported symbols of a library form its API\@.
Given a function $\mathit{exported}: \mathcal{S} \rightarrow \{\top, \bot\}$ indicating whether a given symbol is exported ($\top$) or not ($\bot$), derived from the language's semantics, the API of a library is the set $\mathcal{A} \subseteq \mathcal{S}$ such that $\mathcal{A} = \{s \in \mathcal{S}: \mathit{exported}(s) = \top\}$.
In Java, for instance, visibilities and modules are the primary mechanisms for controlling symbol exports.
As an illustrative example, consider the simplified excerpt of the JDK's \texttt{java.util.ArrayList} depicted in \Cref{lst:arraylist}.
In this example, the set $\mathcal{S}$ holds three symbols:~\ijava{public class ArrayList<E>}, \ijava{public boolean add(E)}, and \ijava{private int size} while $\mathcal{A}$ holds the two first symbols only.

\begin{lstlisting}[language = Java, caption={A simplified excerpt of \ijava{java.util.ArrayList}}, label={lst:arraylist}]
public class ArrayList<E> implements List<E> {
    public boolean add(E e) { !*\etclst*! }
    private int size;
    !*\etclst*!
}
\end{lstlisting}

Depending on its properties, the same kind of symbol may be used in various ways in client code.
Therefore, syntactic usage models employ a function $\mathit{uses}: \mathcal{A} \to \mathcal{P}(\mathcal{U})$, with $\mathcal{U}$ the set of all possible uses, that associates a given symbol with the set of its legal uses.
For instance, following Java's semantics, a \ijava{public class A} can be \emph{Instantiated}, \emph{Referenced}, or \emph{Extended} in client code. In contrast, a \ijava{public abstract class B} can only be \emph{Referenced} or \emph{Extended}, and a \ijava{public class E extends Exception} can even be \emph{Thrown}.
There is no universally correct definition of the $\mathit{uses}$ function:~it is the SUM designer's responsibility to specify the uses of interest that they would like to associate with each individual symbol.
For instance, the use \emph{Referenced} above may be refined for each kind of reference to a type (\eg as a parameter, variable, or return type) if subsequent analyses require it.
Finally, the syntactic usage model $\mathcal{M}$ of a library is the union of the possible uses for every exported symbol in its API $\mathcal{M} = \bigcup_{s \in \mathcal{A}}^{}uses(s)$.
In the example of \Cref{lst:arraylist}, the SUM holds two exported symbols and five possible uses of the API, as shown in \Cref{tab:sum}.

\begin{table}[bth]
    \centering
    \footnotesize
    \caption{Syntactic usage model of \Cref{lst:arraylist}}
    \label{tab:sum}
    \begin{tabular}{rlcp{2.3cm}}
        \textbf{Symbol} & \textbf{Kind} & \textbf{Exported}   & \textbf{Uses} \\
        \midrule
        \ijava{public class ArrayList<E>} & Class & \ding{51} & \parbox{2.2cm}{\raggedright $\{$\textit{Instantiated, Referenced, Extended}$\}$} \\
        \ijava{public boolean add(E e)} & Method & \ding{51} & $\{\mathit{Invoked}, \mathit{Overridden}\}$ \\
        \ijava{private int size} & Field & \ding{55} & $\emptyset$ \\
        \bottomrule
    \end{tabular}
\end{table}

\subsection{Syntactic Usage Footprint (SUF)}
\label{sec:footprint}

While the SUM holds the legal, possible uses of an API, the SUF materializes actual uses of the API in a given piece of client code.
Formally, the footprint of a client on a SUM is the set of triples $\mathcal{F} = \{ \langle s, u, l \rangle : s \in \mathcal{A}, u \in \mathit{uses}(s), l \in \mathcal{L} \}$ where $s$ denotes the API symbol being used, $u$ the kind of use (\eg \textit{Referenced}, \textit{Invoked}), and $l$ the location of the use in client code (\eg a physical line-column location).
As the definition implies, there can be multiple identical uses of the same symbol in different locations.

Syntactic usage footprints materialize the diversity of ways a given API can be used in client code.
For instance, \Cref{lst:lib-snippet} depicts a classical interaction with the \texttt{ArrayList} API through instantiation and invocations.
On the other hand, \Cref{lst:fw-snippet} depicts a typical framework-like interaction with the same API through extension and overriding.
This latter style is prevalent in frameworks such as web servers or batch processing frameworks (\eg Spring, Hadoop, Spark) that heavily employ inversion of control and the Hollywood principle (``\textit{Don't call us, we'll call you}'').
As \Cref{tab:footprint} shows, these two snippets yield two disjoint and complementary SUFs.
Importantly, when looking at multiple clients, their SUFs can trivially be joined through set union and compared through set difference.

\begin{lstlisting}[language = Java, caption={Classical usage of the API of \Cref{lst:arraylist}}, label={lst:lib-snippet}]
ArrayList<Integer> lst = new ArrayList<Integer>();
lst.add(42);
lst.add(1337);
\end{lstlisting}

\begin{lstlisting}[language = Java, caption={Framework-like usage of the API of \Cref{lst:arraylist}}, label={lst:fw-snippet}]
class MyArrayList<E> extends ArrayList<E> {
    @Override
    public boolean add(E e) { !*\etclst*! }
}
\end{lstlisting}

\subsection{Usage Coverage and Metrics}
\label{sec:metrics}

The SUM gives the universe of legal uses to cover for a given API and the SUF pinpoints those that are actually covered in client code.
The set of API symbols that are covered in a SUF is denoted $\mathcal{C}_\mathcal{A} = \{ s : \langle s, u, l \rangle \in \mathcal{F} \}$ and the set of covered API uses is denoted $\mathcal{C}_\mathcal{M} = \{ u : \langle s, u, l \rangle \in \mathcal{F} \}$.
Then, the API symbol coverage score of the SUF with respect to the API is $\frac{\lvert \mathcal{C}_\mathcal{A} \rvert}{\lvert \mathcal{A} \rvert}$ and its API use coverage score is $\frac{\lvert \mathcal{C}_\mathcal{M} \rvert}{\lvert \mathcal{M} \rvert}$.
Following these definitions, \Cref{lst:lib-snippet} covers 50\% of \Cref{lst:arraylist}'s API uses and 100\% of its symbols, so does \Cref{lst:fw-snippet}.
Together, they cover 100\% of the API uses.
One can identify the exported symbols that are not covered in client code by exploring $\mathcal{A} \setminus \mathcal{C}_\mathcal{A}$ and the legal uses that are not realized by exploring $\mathcal{M} \setminus \mathcal{C}_\mathcal{M}$.
Interestingly, one may reproduce the popularity metrics used to identify hotspots and coldspots in the literature~\cite{qiu2016understanding, sawant2017fine} by counting the occurrences of $\langle s, u, l \rangle \in \mathcal{F}$ for a given symbol of interest $s$.

We distinguish between three levels of coverage for each individual API symbol.
An API symbol $s \in \mathcal{A}$ is fully covered if all of its uses are present in the SUF, \ie if $\forall u \in uses(s), \exists \langle s, u, l \rangle \in \mathcal{F}$.
Conversely, an API symbol is not covered if none of its uses are present in the SUF, and it is partially covered if only some of its uses are present in the SUF\@.

%% file: ucov.tex
\section{UCov: Java Syntactic Usage Analysis}
\label{sec:ucov}

\begin{figure}[tb]
	\centering
	\includegraphics[width=\linewidth]{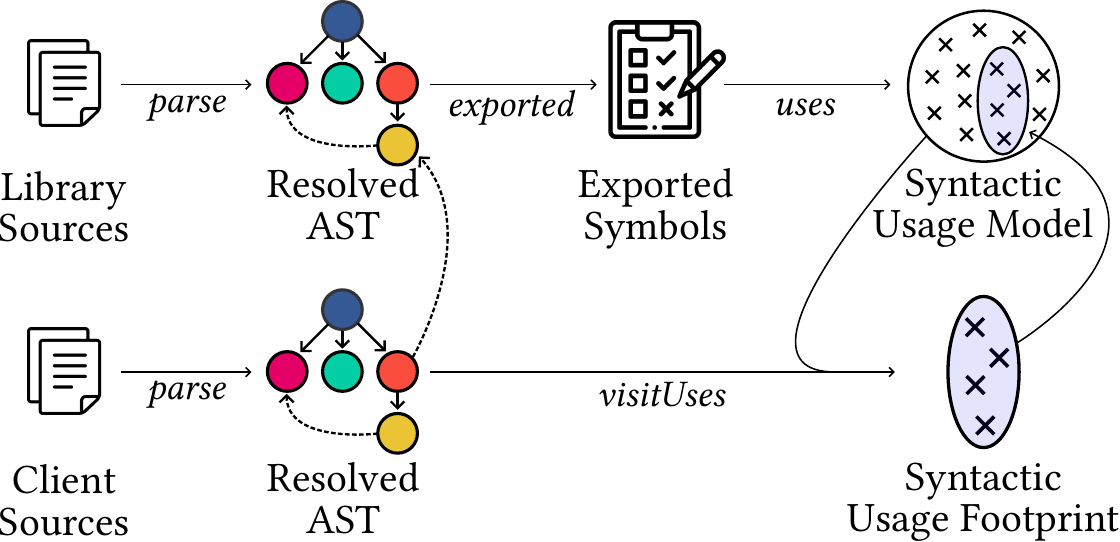}
	\caption{\ucov parses and analyses Java code to build SUM and SUF models. Each cross symbol ($\times$) represents one possible use of an API symbol.}
	\label{fig:ucov}
\end{figure}

\ucov is our proof-of-concept implementation of syntactic API usage analysis for Java libraries.\footnote{UCov is available at \url{https://github.com/alien-tools/ucov}. A snapshot of \ucov alongside all artifacts discussed in this paper is available on Zenodo~\cite{monce_2024_10571867}.}
It parses and analyzes the source code of Java libraries to produce syntactic usage models and the source code of Java clients to produce syntactic usage footprints and their coverage.
When implementing such a tool, one must decide on the symbols of interest to include in the API and the types of uses to represent, based on language semantics and analysis goals, as highlighted in \Cref{sec:models}.
In this section, we discuss some of the choices we made while implementing \ucov to support our case studies and the overall architecture of the tool, depicted in \Cref{fig:ucov}.

\subsection{Exported Symbols}
\label{sec:symbols}
SUM models employ a simple $exported: \mathcal{S} \rightarrow \{\top, \bot\}$ function that indicates whether a given symbol $s$ is exported.
From the library's code, \ucov first extracts the list of all API symbols (namely, for Java, types, methods, and fields) uniquely identified by their fully qualified name (and their signature in the case of methods).
Then, it distinguishes those that can be accessed from the outside, in client code, and those that cannot.
Following the rules described in the Java~17 Language Specification~\cite{gosling2021java}, the following symbols can be accessed from client code:

\begin{description}
    \item[Public symbols:] (transitively) \ijava{public} types, as well as \ijava{public} methods and fields within \ijava{public} types, can be accessed from anywhere without restriction;
    \item[Protected symbols:] \ijava{protected} fields and methods within effectively extensible \ijava{public} types can be accessed through subtyping or by code located in the same package. An effectively extensible type is one that is not \ijava{final} nor \ijava{sealed} and that, in the case of classes, possesses a \ijava{public} or \ijava{protected} constructor the subclass can access;
    \item[Package-private symbols:] package-private symbols (Java's default visibility in the absence of a visibility modifier) can be accessed by code located in the same package.
\end{description}

The only legal visibilities for a top-level type declaration are \ijava{public} and package-private, so API designers can only use the latter when they want to hide a type declaration from the outside.
\ucov's implementation of the $\mathit{exported}$ function thus considers the \ijava{public} and \ijava{protected} symbols and discards the package-private ones, as client code intentionally using a package of the same name as the library's to access its package-private symbols is a pathological case that most likely breaches the API's intent.

\subsection{Supported API Symbol Uses}
\label{sec:ucov-uses}
Once the exported symbols of a library are identified, \ucov maps each of these to a set of legal uses, based on its declaration specifics, thereby implementing the $\mathit{uses}$ function.
\Cref{tab:uses} gives an excerpt of the uses we consider for each kind of symbol.

In a nutshell, \ucov looks for every possible use of type references (as parameter types, return types, exceptions, \etc) and categorizes them all as \emph{Type reference} uses.
Classes can be instantiated if they provide a \ijava{public} (possibly default) constructor and are not \ijava{abstract}.
They can be inherited unless declared \ijava{final}.
Interfaces can be implemented by client classes or extended by other interfaces.
Methods can be invoked (statically or dynamically) and overridden unless declared \ijava{final}.
When polymorphism comes into play in client code and the static analysis cannot determine the dynamically invoked method beyond the static type of its receiver object, \ucov registers an invocation to the corresponding method in the static type of the receiver object, as well as invocations to all super-methods with the same signature in its hierarchy.
As a result, abstract methods in \ijava{abstract} classes and interfaces are considered covered whenever a more concrete implementation is invoked in client code.
For instance, an invocation of \ijava{ArrayList.size()} in client code would cover \ijava{List.size()}.
Finally, \ucov distinguishes between read and write accesses to an exported field.

Note that these choices aim to reflect customary uses of Java APIs and are, to some extent, arbitrary.
Another implementation may, for instance, distinguish between various references to a type or unify read and write accesses to a field under the same kind of use.
These choices will vary depending on the task at hand and analysis objectives.

\begin{table*}[bth]
	\small
	\caption{The kinds of uses considered in \ucov, inspired and refined from the work of \citeauthor{qiu2016understanding}~\cite{qiu2016understanding}}
	\begin{tabular}{llll}
		\textbf{Symbol Type} & \textbf{Use}      & \textbf{Example}                     & \textbf{Resolved API Symbol} \\
		\midrule
		\textbf{Type}        & Reference         & \ijava{String s}                     & \ijava{java.lang.String} \\
                     &                   & \ijava{void f(Integer i)}            & \ijava{java.lang.Integer} \\
                     &                   & \ijava{Integer f()}                  & \ijava{java.lang.Integer} \\
                     &                   & \ijava{void f() throws IOException}  & \ijava{java.io.IOException} \\
                     &                   & \ijava{catch (IOException e)}        & \ijava{java.io.IOException} \\
                     &                   & \ijava{List<String> l}               & \ijava{java.lang.String} \\
                     &                   & \etc                                 & \etc \\
   	\midrule
		\textbf{Class}       & Instantiation     & \ijava{new Integer(42)}              & \ijava{java.lang.Integer} \\
		                     & Inheritance       & \ijava{class T extends Thread}       & \ijava{java.lang.Thread} \\
		\midrule
		\textbf{Interface}   & Implementation    & \ijava{class R implements Runnable}  & \ijava{java.lang.Runnable} \\
  & & \ijava{Runnable r = () -> \{...\}} & \texttt{java.lang.Runnable} \\
		                     & Extension         & \ijava{interface R extends Runnable} & \ijava{java.lang.Runnable} \\
    \midrule
		\textbf{Constructor} & Invocation        & \ijava{new Integer(42)}              & \ijava{java.lang.Integer(int)} \\
		\midrule
		\textbf{Method}      & Invocation        & \ijava{"a".length()}                 & \ijava{java.lang.String.length()} \\
		                     & Static invocation & \ijava{String.valueOf(42)}           & \ijava{java.lang.String.valueOf(int)} \\
		                     & Overriding        & \ijava{@Override void run()}         & \ijava{java.lang.Thread.run()} \\
                       & & \ijava{Runnable r = () -> \{...\}} & \ijava{java.lang.Runnable.run()} \\
		\midrule
		\textbf{Field}       & Field read        & \ijava{Integer.MAX\_VALUE}           & \ijava{java.lang.Integer.MAX\_VALUE} \\
		                     & Field write       & \ijava{Point.x = 2}                  & \ijava{java.awt.Point.x} \\
		\midrule
	\end{tabular}
	\label{tab:uses}
\end{table*}

\subsection{Implementation}
\ucov is implemented in Java and uses the Spoon framework~\cite{pawlak15spoon} to parse Java code and create visitors that implement the necessary static analyses.

When supplied with library code, \ucov uses Spoon to build a resolved AST and applies a dedicated visitor on types, methods, and fields to list the exported API symbols, thereby implementing the $\mathit{exported}$ function (upper part of \Cref{fig:ucov}).
Then, it maps each of these symbols to the corresponding uses ($\times$) according to the rules in \Cref{sec:ucov-uses}, forming the final syntactic usage model.

When supplied with client code (library tests, library samples, code of third-party clients, or any arbitrary snippet using the library), \ucov builds its resolved AST and visits it to identify every node that references, in one way or another, symbols of the API\@.
For example, it analyzes method invocations, field accesses, and type references to determine if they correspond to any of the uses identified in the syntactic usage model.
\Cref{tab:uses} lists an excerpt of the elements in client code that the static analyzer searches for and the information it collects for each.
The resulting set of actual uses in client code constitutes its syntactic usage footprint.

%% file: eval.tex
\section{Exploratory Case Study}
\label{sec:eval}
In this section, we evaluate the capability of syntactic usage models and footprints, as well as their implementation for Java libraries in \ucov, to produce meaningful information that assists maintainers in understanding the boundaries of their libraries and their usage in the wild.
To this end, we follow an exploratory case study approach~\cite{ralph2020empirical} to analyze three popular Java libraries that exhibit diverse interaction styles:~\jsoup, a library for HTML manipulation;~Apache \cli, a library for implementing command-line interfaces; and \spark, a simple web framework.
We introduce our subject libraries in \Cref{sec:subjects} and our methodology in \Cref{sec:methodology}, and then detail our results in \Cref{sec:results}.
We direct the reader to our reproduction package to access the data, scripts, and analyses conducted in this section~\cite{monce_2024_10571867}.


\subsection{Subject Libraries}
\label{sec:subjects}
The implementation of syntactic usage models and footprints in \ucov aims to explore the boundaries of APIs and their actual uses in client code.
Every API is unique, and it is neither possible nor desirable to study all of them.
Instead, we establish a set of criteria to select subject libraries that are diverse and representative of mature Java libraries.

\paragraph{Interaction styles}
Based on our modeling of syntactic usage, we hypothesize that different interaction styles favor different kinds of uses in client code and yield different usage profiles.
To explore the diversity of uses, we aim for libraries that expose diverse styles.
We categorize interaction styles into three types that are widespread in practice:~classical, framework, and fluent. Below, we provide an example of each and explain their significance in the context of API usage.
Naturally, these styles are not mutually exclusive and a library may offer different styles to interact with the same features.

In the \emph{classical} style, APIs expose a set of public classes that are then instantiated in client code to invoke their methods and access their services. Client code retains control of the execution flow. The first snippet of \Cref{lst:styles} depicts a typical classical use of \cli where a parser and options are instantiated and configured using their methods.
The classical style is the most common and is especially popular in libraries that offer a collection of utilities (\eg Google's Guava and Apache's Commons).

In the \emph{framework} style, APIs implement inversion of control and the Hollywood principle to allow client code to extend and specialize types of the API (usually interfaces and abstract classes) by providing implementation code. The library retains control over the execution flow and only hands it to client code when needed. This style is prevalent in frameworks (\eg Spring, Hadoop), hence its name. The second snippet of \Cref{lst:styles} shows a typical framework-like interaction with \spark, where users configure routes for their application by passing lambda expressions that implement the functional interface \ijava{Route} and its unique method \ijava{handle(Request, Response)} which gives the implementation of endpoints.

In the \emph{fluent} style, APIs heavily employ programming tricks such as method chaining and cascading, static methods, the \textit{Builder} design pattern, and static imports to mimic the look and feel of a domain-specific language~\cite{fowler_fluent}.
The third snippet of \Cref{lst:styles} shows a typical fluent interaction with \jsoup that shows how the \textit{Builder} pattern is used to configure a \jsoup connection and load a document from a remote URL\@.

\begin{lstlisting}[language = Java, float, caption={Diverse API interaction styles exemplified using actual documentation samples from our subject libraries}, label={lst:styles}]
// Classical usage of commons-cli
// https://commons.apache.org/proper/commons-cli/usage.html
CommandLineParser parser = new DefaultParser();
Options options = new Options();
options.addOption("a", "all", false, "do not hide entries");
options.addOption("C", false, "list entries by columns");
try {
	CommandLine line = parser.parse(options, args);
	if (line.hasOption("block-size")) { !*\etclst*! }
}
catch (ParseException exp) { !*\etclst*! }

// Framework-like usage of Spark
// https://sparkjava.com/documentation#routes
get("/",     (request, response) -> { !*\etclst*! });
post("/",    (request, response) -> { !*\etclst*! });
put("/",     (request, response) -> { !*\etclst*! });
delete("/",  (request, response) -> { !*\etclst*! });
options("/", (request, response) -> { !*\etclst*! });

// Fluent-like usage of jsoup
// https://jsoup.org/cookbook/input/load-document-from-url
Document doc = Jsoup.connect("http://example.com")
	.data("query", "Java").userAgent("Mozilla")
	.cookie("auth", "token").timeout(3000).post();
\end{lstlisting}

\paragraph{Client code}
To make the analysis possible, we look for libraries that have sufficient amounts of client code available:~third-party clients, tests, and documentation samples.
This rules out immature libraries that lack sufficient documentation or clients, or that are not well-tested.

While we expect to find third-party clients and tests that can be parsed and analyzed with \ucov, documentation and samples suffer from additional issues.
Some libraries include their samples as proper compilable files in their source directory, while others include their samples in readme files (\eg on GitHub) or dedicated documentation websites.
These samples are typically simple snippets cleaned from the surrounding boilerplate code (imports, structure, type declarations, \etc) that often cannot be parsed on their own.
When these cases arise, we create a Java file for each case, encapsulating the snippet within a main method and manually inserting the missing imports.
For our subject libraries, with the help of the surrounding documentation, the missing imports are always unambiguous.

\paragraph{Selected libraries}
To obtain high-quality libraries that meet these requirements, we start from the Duets dataset~\cite{durieux2021duets}.
Duets contains 395 libraries and 2,874 clients extracted from GitHub that compile, can be executed, have passing test suites and a minimum of five stars.
To identify libraries with documentation samples, we narrow our search to libraries with a readme file and an official documentation website.
If these contain documentation samples, we collect them. Otherwise, we follow their hyperlinks and repeat the process.
We also explore the source tree of each library to find samples stored alongside the library code.
Finally, we handpick a set of three libraries that exhibit diverse interaction styles and have sufficient test cases, documentation samples, and third-party clients: \jsoup (classical and fluent styles), \cli (classical and fluent styles), and \spark (framework style).
\Cref{tab:descriptive-statistics} provides some descriptive statistics for these libraries.

For these three libraries, we retrieve the following documentation samples:
\begin{itemize}
    \item For \cli, its readme links to the official documentation, which includes a page showcasing some sample uses.\footnote{\url{https://commons.apache.org/proper/commons-cli/usage.html}}
    This page contains partial code snippets without imports or structure.
    We manually reconstruct parseable Java files for these snippets.
    \item For \jsoup, we find a set of samples directly in its source code (\ijava{src/main/java/org/jsoup/examples}). Its readme file also leads to the official website and a cookbook that presents different sample snippets.\footnote{\url{https://jsoup.org/cookbook/}}
    We reconstruct complete Java files with proper imports and structure for these samples.
    \item For \spark, its readme includes fully parseable sample code and links to the official documentation.\footnote{\url{http://sparkjava.com/documentation}}
    The documentation points to a list of user-authored tutorials,\footnote{\url{https://sparkjava.com/tutorials}} which consist of parseable Java files, and a list of template projects,\footnote{\url{http://sparkjava.com/tutorials/application-structure}, \url{https://github.com/tipsy/spark-basic-structure}, \url{https://github.com/perwendel/spark-template-engines}} which we also include.
\end{itemize}

\begin{table}[tb]
    \caption{Descriptive statistics of the subject libraries, extracted from GitHub on October 30, 2023}
    \begin{tabular}{lrrr}
                     & \cli       & \jsoup     & \spark \\
       \midrule
        Last release date & 2021/10/29 & 2023/04/29 & 2020/10/08 \\
        Version      & 1.5.0      & 1.16.1     & 2.9.3 \\
        Since        & 2002       & 2010       & 2011 \\
        Stars        & 309        & 10.4k      & 9.5k \\
        Commits      & 1,374      & 1,889      & 1,067 \\
        Size (LoC)   & 6,307      & 27,817     & 11,298 \\
        Contributors & 46         & 97         & 100 \\
        Clients      & 49k        & 131k       & 30k \\
        \bottomrule
    \end{tabular}
    \label{tab:descriptive-statistics}
\end{table}

\subsection{Methodology}
\label{sec:methodology}

Our methodology aims to explore the interactions offered by various APIs and their actual uses in third-party clients, tests, and documentation samples.
Since the number of clients per library stored in Duets is relatively low ($31$ for \jsoup, $202$ for \cli, and $8$ for \spark), and as we do not require clients to have passing test suites, we collect new clients for each of the libraries using GitHub's dependency graph.\footnote{\url{https://docs.github.com/en/code-security/supply-chain-security/understanding-your-software-supply-chain/about-the-dependency-graph}}
We compile a list of every repository in the dependency graph that declares a dependency towards one of the libraries of interest.
After discarding repositories that cannot be retrieved and forks, we obtain 11,159 clients for \jsoup, 11,612 for \cli, and 12,530 for \spark.
Naturally, it is not necessary to analyze all of these, so we apply the standard Cochran formula with a confidence level of $c = 95\%$, an error margin of $e = 5\%$, and a conservative proportion of $p = 0.5$.
We obtain sample sizes of 372, 372, and 373, which we draw at random from the corresponding client sets.
These sets contain the third-party clients, which we analyze in the remainder of this section.

We use \ucov to construct the SUM of each library.
Following the specification in \Cref{sec:ucov}, \ucov extracts the list of all exported API symbols and maps them to their corresponding legal uses.
The resulting models materialize the universe of interactions enabled by the libraries (\Cref{tab:lib-sums}).
For each library, we classify client code as either tests, documentation samples, or third-party clients, and we run \ucov on these three corpora to produce their SUF\@.
Since SUFs can be easily joined through set union, we also construct a merged SUF for each library representing every use from all client code, denoted \textit{All} in \Cref{tab:lib-sums}.

\begin{figure*}[tb]
    \centering
    \begin{subfigure}{.49\linewidth}
        \includegraphics[width=\linewidth]{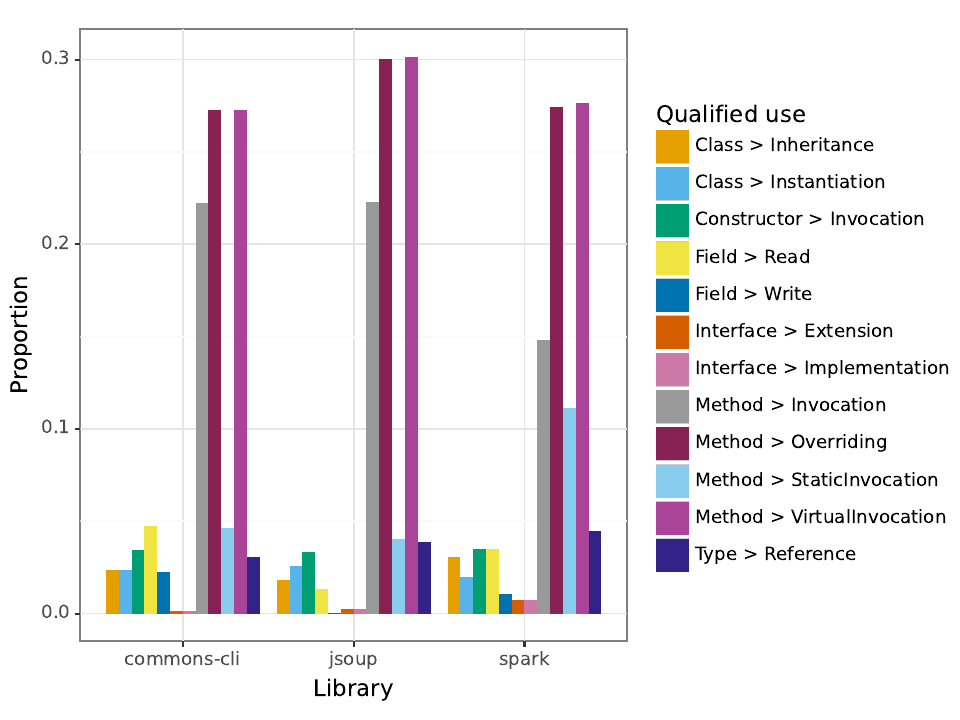}
        \caption{As a distribution of the legal uses in their APIs (SUM)}
        \label{fig:sum-profiles}
    \end{subfigure}
    \begin{subfigure}{.49\linewidth}
        \includegraphics[width=\linewidth]{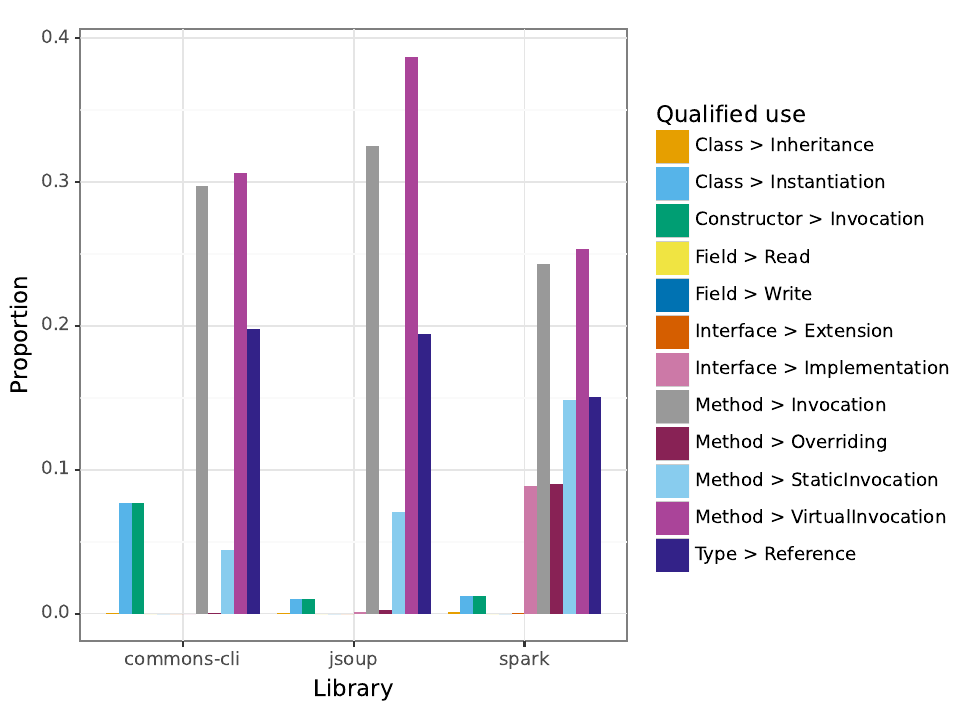}
        \caption{As a distribution of the actual uses in client code (SUF)}
        \label{fig:suf-profiles}
    \end{subfigure}
    \caption{Usage profiles of the three libraries \cli, \jsoup, and \spark}
\end{figure*}

\subsection{Analysis and Results}
\label{sec:results}

This section discusses the results obtained with \ucov for the three subject libraries and their client code.

\begin{table}[tb]
    \centering
    \small
    \caption{Syntactic usage models extracted from \cli, \jsoup, and \spark, and syntactic usage footprints extracted from their third-party clients, tests, and samples. Symbols used are presented as \% of API symbols and unique uses as \% of legal uses.}
    \begin{tabular}{p{.4cm}p{1cm}lrrr}
     &  &  &  \textsc{cli} &\jsoup & \spark \\
     \midrule
    \multicolumn{2}{l}{\multirow{2}{*}{\textbf{SUM}}} & API symbols &  291 &1,138 & 771 \\
     &  & Legal uses &  755 &2,941 & 1,960 \\
     \midrule
    \multirow{12}{*}{\textbf{SUF}} & \multirow{4}{*}{\parbox{1cm}{Symbols used}} & All &  205 (70\%)&608 (53\%)& 301 (39\%)\\
     &  & Clients &  153 (53\%)&282 (25\%)& 179 (23\%)\\
     &  & Tests &  179 (62\%)&586 (51\%)& 248 (32\%)\\
     &  & Samples &  24 (~8\%)&47 (~4\%)& 134 (17\%)\\
      \cmidrule(lr){2-6}
     & \multirow{4}{*}{\parbox{1cm}{Unique uses}} & All &  367 (49\%)&1,028 (35\%)& 476 (24\%)\\
     &  & Clients &  266 (35\%)&456 (16\%)& 278 (14\%)\\
     &  & Tests &  315 (42\%)&967 (33\%)& 400 (20\%)\\
     &  & Samples &  38 (5\%)&69 (~2\%)& 211 (11\%)\\
     \cmidrule(lr){2-6}
     & \multirow{4}{*}{\parbox{1cm}{Total uses}} & All &  79,284&29,979& 23,882\\
     &  & Clients &  74,453&15,246& 20,827\\
     &  & Tests &  4,690&14,511& 1,605\\
     &  & Samples &  141 &222 & 1450\\
    \bottomrule
    \end{tabular}
    \label{tab:lib-sums}
\end{table}

\paragraph{Syntactic usage models and footprints}
\Cref{tab:lib-sums} presents simple statistics for the SUMs and SUFs extracted from our three subject libraries.
\jsoup exposes the most extensive API, with 1,138 symbols and 2,941 legal uses in its SUM, followed by \spark and \cli.
This is consistent with their overall size in lines of code (\Cref{tab:descriptive-statistics}).
The three libraries exhibit a similar ratio of exported API symbols to legal uses, averaging around 2.5 legal uses per symbol.

The coverage scores for uses are consistently lower than for API symbols.
This is expected, as covering a use implies covering the corresponding symbol.
However, the sizeable difference in coverage indicates that many of the interactions legally allowed in the APIs are not realized in either tests, documentation samples, or third-party clients, even when the corresponding symbols are known and used.
Although the relatively low coverage of API symbols in client code has been extensively studied in the literature (\eg~\cite{harrand2022api,qiu2016understanding}), this suggests that symbol coverage does not fully reflect the extent of interactions permitted by APIs.

Across all libraries, tests achieve the highest coverage scores for symbols and uses, followed by third-party clients and samples.
This suggests that the tests cover a sizeable subset of the APIs.
According to Hyrum's law, ``\textit{with a sufficient number of users of an API \textelp{} all observable behaviors \textelp{} will be depended on by somebody}''~\cite{hyrum}, suggesting that third-party clients might eventually achieve the highest coverage score, assuming a sufficient number of clients. This has already been empirically verified for the libraries hosted in Maven Central~\cite{harrand2022api}.
Documentation samples are sparse, so naturally they cover much less than tests and third-party clients.
\spark stands out with a much better coverage score of its API by samples, thanks to the rich documentation on its official website.

Overall, \cli is the most focused library, with fewer exported symbols and possible uses, which results in better coverage of its symbols and uses in client code.
Its API also exhibits repetitive patterns:~client code often instantiates numerous command-line options and configures them similarly, resulting in frequent and uniform usage of identical symbols (averaging 216 instances per use, compared to 29 for \jsoup and 50 for \spark).

\paragraph{Library profiles}
\Cref{fig:sum-profiles} depicts the usage profiles of our three libraries.
The profile of a library is the distribution of the legal uses in their SUMs, such that the proportion of each kind of use adds up to 1.
It highlights which kinds of uses are allowed and which are the most frequent.
The profiles are mostly similar, with some specificities for each library.
In all cases, (virtual) method invocation and method overriding dominate the frequencies.
This is expected as methods are by far the most frequent kind of symbol in the three libraries, with fewer types and fields exposed.
\cli and \jsoup share a very similar profile.
Indeed, they both implement the classical and fluent styles in their interactions.
Interestingly, \spark exposes a larger number of methods that can be statically invoked and interfaces that can be extended and overridden in client code.
These correspond to the primary interfaces and methods used in client code to configure routes (\ijava{get()}, \ijava{post()}, \ijava{path()}, \etc), as shown in \Cref{lst:styles}.
This is consistent with its framework-like interaction style.

\paragraph{Usage analysis}
\Cref{fig:suf-profiles} shows the usage profiles of the three libraries, defined as the distribution of actual uses in their SUFs (including third-party clients, tests, and samples).
Comparing \Cref{fig:sum-profiles} and \Cref{fig:suf-profiles} reveals the discrepancies between what the APIs allow and what client code actually uses.

Among the three libraries, method overriding is the least covered type of use by a large margin.
This suggests that the APIs offer possibilities for extension and specialization that are not yet utilized in client code or, more likely, that many of the API methods could be closed for extension using the \ijava{final} keyword at the level of the method or its containing type.
Until version 17 (2021) and the introduction of sealed classes, Java lacked the ability to restrict extension and overriding to a predetermined set of types.
Without this possibility, library maintainers had to open types for extension and specialization to everyone, even when their intent was to allow extension and specialization in library code only.

Here again, \cli and \jsoup expose a similar profile.
However, \jsoup's clients rely more on its fluent style, with fewer class instantiations and more frequent use of static invocations and builders to create objects.
We also observe that although \cli and \jsoup expose some fields that can be read and written, as well as some types that can be extended or implemented, clients rarely use these interactions.
The only uses of these interfaces in client code are through type references.

\spark, on the other hand, exhibits a different profile with many static invocations, method overrides, and interface implementations.
Indeed, as advocated in its documentation, the preferred method of declaring routes in \spark is to pass a lambda expression that implements a single method in the \ijava{Route} interface, resulting in each route declaration involving one static invocation, one interface implementation, and one method override.

Digging deeper into the most popular uses for each library, we find that the SUFs accurately reflect the expected uses of each library, as documented in their samples.
For \cli, the top three most popular interactions involve referencing and instantiating the symbols \ijava{Option} and \ijava{Options}, and invoking the method \ijava{addOption}, which developers use to build their command-line interface.
For \jsoup, the top three most popular interactions involve referencing the symbols \ijava{Document}, \ijava{Element}, and invoking the method \ijava{Element.select(String)}, which developers use to navigate HTML documents.
For \spark, the top three most popular interactions involve implementing the interface \ijava{Route} and overriding its method \ijava{Route.handle()}, which developers use to declare routes, as well as referencing the types \ijava{Request} and \ijava{Response}, which are passed to and from the routes.
We refer the reader to the reproduction package for a comprehensive list of the most and least popular API symbols and interactions.

Interestingly, we observe that \jsoup exposes parts of its \textit{internal} API publicly (\texttt{org.jsoup.internal}), most likely for technical reasons:~isolating types in a package forces them to be \ijava{public} to allow other packages of the library to access them.
While the maintainers take great care to discourage clients from using these APIs using source code comments (``\textit{Jsoup internal use only, please don't depend on this API.}``), we find several uses in the code of third-party clients.
Perhaps more surprisingly, \jsoup's official documentation also uses these internal APIs in one of its samples.\footnote{\url{https://github.com/jhy/jsoup/blob/84fd43766992401057b73f740acec4b82f1e3dd6/src/main/java/org/jsoup/examples/HtmlToPlainText.java}}
\ucov allows library maintainers to identify these problematic cases at a glance.

\begin{figure*}[tb]
	\centering
	\begin{subfigure}{.315\linewidth}
		\includegraphics[width=\linewidth]{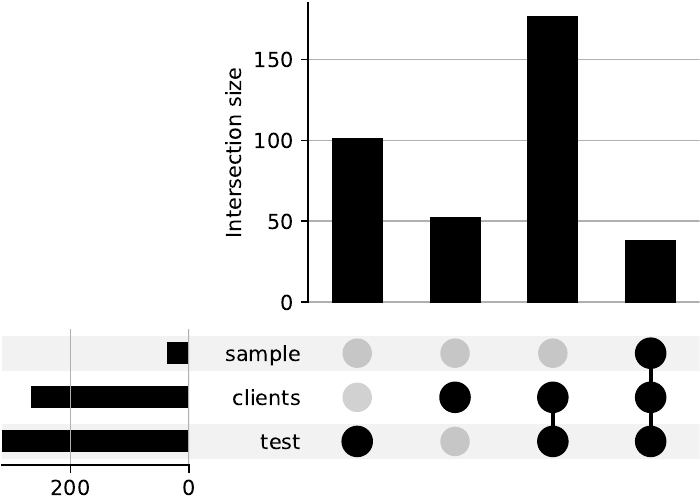}
		\caption{SUF intersections for \cli}
		\label{fig:upset-cli}
	\end{subfigure}
	\quad
	\begin{subfigure}{.315\linewidth}
		\includegraphics[width=\linewidth]{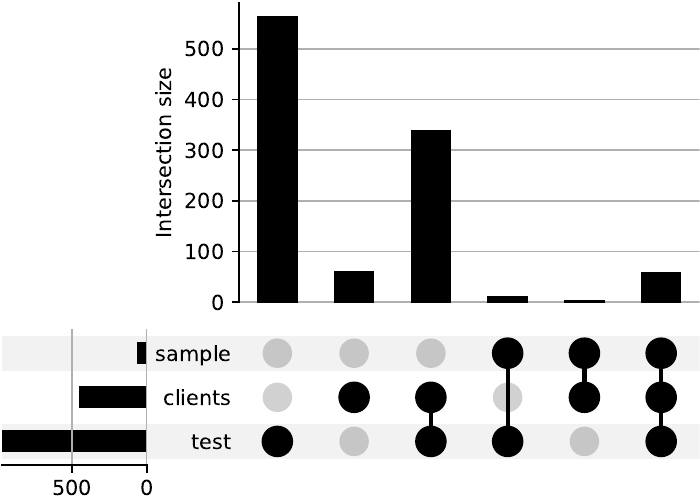}
		\caption{SUF intersections for \jsoup}
		\label{fig:upset-jsoup}
	\end{subfigure}
	\quad
	\begin{subfigure}{.315\linewidth}
		\includegraphics[width=\linewidth]{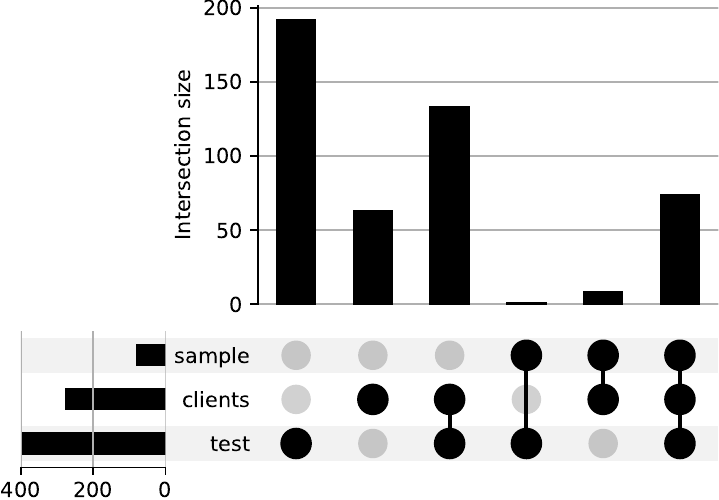}
		\caption{SUF intersections for \spark}
		\label{fig:upset-spark}
	\end{subfigure}
	\caption{UpSet plots~\cite{2014_infovis_upset} depicting the common and unique uses between third-party clients, tests, and samples for \jsoup, \cli, and \spark}
	\label{fig:upset}
\end{figure*}

\paragraph{Comparative analysis of third-party clients, tests, and samples}
The SUF profiles of third-party clients, tests, and samples (not shown here for conciseness but available in the reproduction package) are very similar for each library.
However, when looking at individual symbols and uses, we observe that each has its specificities.
\Cref{fig:upset} depicts the individual contributions to unique uses of third-party clients, tests, and samples, represented as UpSet plots to visualize the size of their intersections.
This visualization enables library maintainers to immediately identify gaps in the coverage of client code by their tests and documentation.

The official samples of \cli\footnote{\url{https://commons.apache.org/proper/commons-cli/usage.html}} are particularly sparse compared to \jsoup and \spark:~all the uses they document are also found in tests and third-party code.
Conversely, we identify 176 uses that are common to tests and third-party clients but not documented (23\% of legal uses).
These include the deprecated \ijava{OptionBuilder} type, which is unsurprisingly not documented, but also symbols that are popular in client code such as \ijava{org.apache.commons.cli.Option.isRequired()}.
As the API of \cli is well-focused, there is still a significant amount of uses that are common to all three.
However, we observe that 52 uses found in third-party clients are not exercised:~for instance, the method \ijava{org.apache.commons.cli.Option.setValueSeparator(char)} is neither tested nor documented.

The samples of \jsoup are more varied, but there is still a significant amount of uses that are undocumented.
Among these 338 uses, we identify some popular interactions such as invoking the \ijava{org.jsoup.nodes.Node.hasAttr(java.lang.String)} method.
Its tests are comprehensive, and only a few uses are exclusive to third-party clients.
Among these, we find clients extending and overriding the \ijava{org.jsoup.Connection} type to implement their own logic and innocuous cases such as invocations to the many \ijava{toString()} methods that are (predictably) untested and undocumented.
Interestingly, some of the uses we find in third-party clients are also documented in its samples, but absent from the tests.
This is, for instance, the case for the internal APIs of \jsoup discussed previously.

Despite the extensive samples found in its readme and tutorials, \spark still presents 39 uses that are common to clients and tests but undocumented.
Even more surprisingly, we find some uses that are documented but neither tested nor used in third-party clients, such as invocations of the \ijava{spark.Request.bodyAsBytes()} method.
In contrast with \cli and \jsoup, we find some uses that are documented, tested, but for which we could not find any instance in client code:~we did not find any static invocation of the method \ijava{spark.Spark.modelAndView(Object, String)} which is the documented way of instantiating a \ijava{ModelAndView} object.
Similarly to \jsoup, we find 27 API interactions that are documented in samples and used by third-party clients but remain untested.
These include setting the listening port of the web server (\ijava{spark.Spark.port(int)}) and examining the underlying raw Java requests beneath \spark's \ijava{Request} objects (\ijava{spark.Request.raw()}).

A significant part of the value of syntactic models resides in the intersection of the uses of different kinds of client code.
\ucov allows library maintainers to identify undocumented or untested interactions in their APIs and take appropriate action. Naturally, it is not practical to expect that all API interactions are covered by the library's tests or samples.
Most developers use IDEs to discover features, and the associated JavaDoc is often sufficient.
Besides, trivial APIs such as the \ijava{toString()} methods mentioned above may not necessarily require documentation.
Nevertheless, we believe that \ucov can help maintainers prioritize their documentation and testing efforts by focusing on the interactions that are commonly used in the code of third-party clients, and identify blind spots in their design.

%% file: discussion.tex
\section{Discussion}
\label{sec:discussion}
Syntactic usage models and syntactic usage footprints offer a lightweight and intuitive way to analyze the interactions allowed by an API and their coverage in client code.
While our case study primarily explores their ability to accurately represent different library profiles and identify unused, undocumented, and untested interactions, they can inform and benefit other scenarios.

\paragraph{Support for API design in programming languages}
SUM models represent the interactions permitted by an API\@.
When using them, it becomes clear that the capabilities provided by Java for designing APIs are limited.
Library developers are often required to expose numerous legal interactions to client code for purely technical reasons that have little to do with API design.
For example, although library developers can prevent extension and overriding for everyone by using the \ijava{final} modifier, they have only recently gained the ability to open extension to a predetermined set of implementers using \ijava{sealed} classes.
On the other hand, it is not possible to fine-tune the interactions allowed in client code, such as allowing method overriding but not method invocation, which can violate the principles of inversion of control in framework-like libraries.
Similarly, there are no mechanisms to restrict access to certain types, fields, and methods beyond the basic scoping mechanisms provided by visibilities.
As a result, developers must sometimes declare types as package-private or even public for technical reasons and rely on code comments and naming conventions such as internal packages and \ijava{@Internal} annotations to warn their users.
SUM models make all these issues explicit and could facilitate reflection on language design when adapted to other programming languages.

\paragraph{Compatibility and breaking changes}
When libraries evolve, they sometimes introduce breaking changes that affect clients and force them to adapt their code.
How breaking changes impact client code depends on the interactions with the broken API\@.
For example, a class that evolves into an abstract class does not break client code that references it but does break client code that attempts to instantiate it.
Simply knowing which API symbol the client code uses is insufficient to assess the impact of breaking changes.
Some approaches attempt to measure the impact of potential breaking changes by analyzing client code.
While some only rely on import declarations to determine if client code may be impacted~\cite{xavier2017historical}, more recent approaches gather usage information to improve the accuracy of impact detection~\cite{ochoa2022breaking,ochoa22breakbot}.
SUF models provide accurate information regarding the uses made in client code and we believe they could improve the accuracy of these tools.

Another interesting direction is to utilize SUM models to automatically generate synthetic client code that thoroughly exercises every possible interaction with the API\@.
This code would exhibit a complete footprint of the API and could be recompiled whenever the library is updated, allowing the compiler to automatically check if any of the interactions break, signaling a breaking change.
Using the compiler as ground truth for breaking change detection would address the accuracy issues of existing detection tools~\cite{jezek2017api} and automatically cope with the evolution of programming language specifications and their implementation in compilers and virtual machines.

\paragraph{API evolution}
Predicting the consequences of API changes can be challenging.
When an API evolves, it can not only introduce breaking changes but also alter the way clients interact with it, even when the changes are backward-compatible.
Because SUM models can be efficiently computed (e.g., during code review or continuous integration), maintainers can reason about the impact of their changes and refactorings by exploring the differences between pre- and post-change SUMs.
This could allow maintainers to easily assess the impact of external contributions on the interactions permitted by their API and determine whether to incorporate the changes.

%% file: rw.tex
\section{Related Work}
\label{sec:rw}
In this section, we discuss related work on the analysis and applications of API usage.
Specifically, we review the tools and studies that analyze usage at the symbol level, the extensive literature on usage patterns and protocols, and their applications to API evolution and breaking changes analysis.

\paragraph{API symbols analysis}
Traditionally, the analysis of API usage has primarily been conducted at the level of individual symbols.
Many studies analyze how API symbols are used to identify notable \emph{hotspots} and \emph{coldspots}, \ie parts of the APIs that are over- or under-utilized~\cite{stylos2009improving,sawant2017fine,thummalapenta2008spotweb}.
A recent study by \citeauthor{qiu2016understanding} delves into the usage of Java's standard library and third-party libraries across a corpus of over 5,000 projects, encompassing 150M+ lines of code, and finds that usage follows a Zipf distribution~\cite{qiu2016understanding}.
Another study by \citeauthor{harrand2022api} examines 2.2M dependencies on Maven Central and confirms the observations of Hyrum's law:~a small subset of the API attracts most usages in client code~\cite{harrand2022api}.
As illustrated by our exploratory case study, the coverage of uses is much smaller than that of symbols.
Various tools and studies employ different methodologies for collecting usage data, whether from bytecode~\cite{harrand2022api}, source code and resolved ASTs~\cite{qiu2016understanding,lammel2011large, deroover2013multi}, or other sources~\cite{stylos2009improving}.
A unifying theme across these works and their underlying models of API usage is their focus on determining whether a specific API symbol is accessed in client code, rather than exploring how it is used.
Syntactic models and \ucov go beyond this level of granularity by distinguishing the various interactions allowed by individual symbols.
As our case study shows, the coverage of uses in client code is much smaller than that of symbols.

\paragraph{Usage patterns and protocols}
A variety of work in the literature addresses the problem of mining and recommending unordered or sequential API usage patterns and protocols to users~\cite{nguyen19focus,robillard_automated_2013,zhong2009mapo,wang2013mining,buse2012synthesizing}.
These studies typically analyze library code and client code to infer automata and probabilistic graphs that represent legal sequences of API invocations and check whether client code complies with them.
Some even gather data from Q\&A websites to infer and recommend usage patterns~\cite{zhang2018code,uddin2020mining}, while others empirically study these usages in the wild~\cite{zhong_empirical_2017}.
These approaches consider sequences of API calls, often with temporal properties, and go beyond the goals of \ucov.
On the other hand, \ucov still provides a more detailed view of interactions with API symbols, which could benefit these approaches.
In our work, we leverage syntactic usage models and footprints to help developers understand which API uses are allowed and to help them improve their API design, eliminating unintended uses or documenting undocumented areas of their APIs.
A promising area for future research is studying how syntactic usage models and footprints can contribute to extracting more fine-grained usage patterns and protocols.

\paragraph{API evolution breaking changes}
There is a wide range of literature on the nature and driving forces of API evolution~\cite{ketkar2020understanding,kula2018empirical,mcdonnell2013empirical}, as well as on breaking changes and their impact~\cite{ochoa2022breaking,xavier2017historical,brito2018and}.
Some studies focus on detecting breaking changes through API usage patterns~\cite{gao_apifix_2021,zaitsev2022how,lamothe_systematic_2022,zhang_has_2022,venturini_i_2023,nguyen_graph-based_2019,foo_efficient_2018,harrand2022api}.
A key enabler for these studies is the use of accurate models that can represent both the interactions allowed by an API and their uses in client code.
The impact of a breaking change, for example, depends on both the symbol being used and the way it is used.
We believe that the rich information provided by syntactic usage models and footprints can better inform breaking change detection tools~\cite{jezek2017api} and improve their accuracy in assessing the impact of breaking changes on client code. This information can also better inform studies that aim to understand API usage and evolution by refining the level of granularity at which APIs are considered~\cite{businge2015eclipse, businge2019stable}.

%% file: conclusion.tex
\section{Conclusion}
In this paper, we introduced syntactic usage models and footprints to support library maintainers in understanding the boundaries of their APIs and the interactions they allow.
We presented an implementation of these models for the Java programming language in \ucov, a static analysis tool that analyzes the source code of libraries to collect their exported symbols and allowed uses, and client code to collect fine-grained uses of the library.

Our exploratory case study of three popular Java libraries that exhibit diverse interaction styles (\cli, \jsoup, and \spark) showed that \ucov provides valuable information regarding the usage profiles of libraries and their uses in documentation samples, tests, and third-party clients.
Specifically, we showed how \ucov can pinpoint undocumented and untested interactions, as well as misalignments between legal uses of the API and actual uses in client code.

For future work, we will explore the suitability of syntactic usage models and footprints to support graceful API evolution and the detection of breaking changes and their impact.
We will also employ \ucov for large-scale empirical analysis of the evolution of Java libraries and investigate the benefits of syntactic models in other programming languages and ecosystems.